\documentclass[fleqn,usenatbib]{mnras}

\usepackage{newtxtext,newtxmath}

\usepackage[T1]{fontenc}

\let\vec\mathbf
\DeclareRobustCommand{\VAN}[3]{#2}
\let\VANthebibliography\thebibliography
\def\thebibliography{\DeclareRobustCommand{\VAN}[3]{##3}\VANthebibliography}

\usepackage{graphicx}	
\usepackage{amsmath}	

\title[An efficient framework for halo bias modeling]{Stratonovich approximation for halo bias: An efficient framework within excursion set theory}

\author[M. Noorikuhani and S. Baghram]{
Milad Noorikuhani,$^{1, 2}$\thanks{E-mail: m.noorikuhani92@sharif.edu}
Shant Baghram,$^{1, 2}$\thanks{E-mail: baghram@sharif.edu}
\\
$^{1}$Department of Physics, Sharif University of Technology, P.O.Box 11155-9161, Tehran, Iran\\
$^{2}$Research Center for High Energy Physics, Department of Physics, Sharif University of Technology, Tehran 11155-9161, Iran
}

\date{Accepted XXX. Received YYY; in original form ZZZ}

\pubyear{\the\year{}}

\begin{document}
\label{firstpage}
\pagerange{\pageref{firstpage}--\pageref{lastpage}}
\maketitle

\begin{abstract}
We develop an efficient framework for calculating halo multiplicity functions and bias parameters in the excursion-set approach based on the previously presented Hertz and Stratonovich approximations for the first-crossing distribution. The framework enables efficient evaluation of first-crossing distributions for a broad class of moving barriers and filter functions, from which halo biases can be calculated within the peak-background split formalism. We compare halo multiplicity functions and biases obtained from these approximations with those computed from first-crossing distributions obtained from numerically generated ensembles of excursion-set trajectories. We find close agreement between predictions based on the Stratonovich approximation and numerical results across a range of barrier parameters and filter choices, demonstrating that the approximation captures the impact of non-Markovian correlations on first-crossing statistics while substantially reducing the computational cost compared to trajectory-based calculations. Overall, this framework provides a flexible and computationally efficient tool for analyzing halo statistics in cosmological applications. 
\end{abstract}




\section{Introduction}
\label{sec:intro}
The wealth of observational data from modern galaxy surveys provides unique opportunities for testing different cosmological models and identifying those that best describe our universe. For this purpose,  galaxy-halo connection models (e.g. \cite{2018ARA&A..56..435W}) can be used to connect the observed properties and clustering of galaxies to the properties and spatial distribution of their host dark matter halos, whose abundance and clustering are modeled within an underlying cosmological framework. Commonly used methods model the abundance and clustering of dark matter halos at a given redshift in terms of the (statistical) properties of the dark matter density fluctuation field that is linearly evolved from an initial state (e.g. \cite{1974ApJ...187..425P,  1986ApJ...304...15B, 1990MNRAS.243..133P, 1991ApJ...379..440B, 1996MNRAS.282..347M, 1999MNRAS.308..119S, 2002MNRAS.329...61S, 2007IJMPD..16..763Z,  2010ApJ...711..907M, 2010ApJ...717..515M, 2011MNRAS.411.2644M,  2011PhRvL.106x1302C, 2011PhRvD..84b3009C,  2012MNRAS.426.2789P, 2013MNRAS.431.1503P, 2014JCAP...07..044L, 2016arXiv161103619C, 2017PhRvD..96d3524N}). In this context, the cosmological model dependence enters, in part, through the statistics of the initial density fluctuation field and its evolution. 

One of the most widely used frameworks among such approaches is the excursion set theory (EST) (\cite{1991ApJ...379..440B, 1993MNRAS.262..627L}). The basic implementation of the EST involves calculating the smoothed linear density contrast, $\delta(R)$, at different smoothing scales $R$, around a given point in Lagrangian space. Since there is a monotonic relationship between $R$ and the variance of the smoothed density contrast, $S \equiv \langle \delta(R)^2 \rangle$, it is common to map $\delta(R)$ to $\delta(S)$ for calculations. In this manner, as one recursively reduces the smoothing scale $R$ (or equivalently increases $S$), the corresponding smoothed density contrast traces out a ``trajectory'' in the $S-\delta(S)$ plane. This process can then be repeated for numerous other randomly chosen points in the Lagrangian space, creating an ensemble of trajectories which can then be analyzed as a stochastic process. 

In the EST formalism, the first up-crossing of a positive threshold value (the collapse barrier) by a trajectory at a given scale $S$, corresponding to a smoothing scale $R(S)$, is associated with the collapse of the matter enclosed within the corresponding Lagrangian volume into a virialized halo in Eulerian space. The mass of the Eulerian halo can then be related to the mass of the Lagrangian ``protohalo'' region, $M(R)$, which is determined by the smoothing scale $R(S)$ and the filter function used to smooth the density contrast. The simplest collapse barrier is a constant barrier, $B = \delta_c$, motivated by the spherical collapse model (\cite{1974ApJ...187..425P, 1991ApJ...379..440B}). A more physically motivated barrier is the Sheth-Mo-Tormen (SMT) (\cite{2001MNRAS.323....1S}) moving barrier, which depends on $S$ and is motivated by an ellipsoidal collapse model that accounts for tidal effects. There are also other types of barriers, such as stochastic barriers that partially incorporate the complex nature of halo formation by introducing variability/scatter in the collapse threshold (\cite{2010ApJ...717..515M}).
 
The first crossing distribution (FCD), $f(S)$, is therefore a key function in the EST analyses as $f(S) \mathrm{d}S$  determines the fraction of trajectories that up-cross a given barrier for the first time within the interval $(S, S+\mathrm{d}S)$. Within the EST framework, this fraction can be interpreted as the fraction of matter that has collapsed into halos of the corresponding mass interval, obtained by mapping $S$ to the halo mass scale $M$. It is clear that halo abundance quantities, such as the halo mass function which describes the comoving number density of halos per unit mass, are proportional to $f(S)$ and can be calculated from that (e.g. \cite{1991ApJ...379..440B}). 

Another key quantity in cosmological studies is halo bias, which characterizes the clustering of halos relative to the underlying matter density field (e.g. \cite{1984ApJ...284L...9K, 1989MNRAS.237.1127C, 1996MNRAS.282..347M, 1999MNRAS.308..119S, 2010ApJ...724..878T, 2011PhRvD..84b3009C, 2012MNRAS.426.2789P, 2016PhRvD..93f3007L, 2016JCAP...02..018L, 2018PhR...733....1D, 2024A&A...691A..62E}). In general, halo bias also exhibits scale dependence and non-locality, particularly beyond the linear regime (e.g. \cite{1998MNRAS.297..692C,  2007PhRvD..75f3512S, 2009JCAP...08..020M, 2011PhRvD..83h3518M, 2012PhRvD..86h3540B, 2012PhRvD..85h3509C, 2012MNRAS.419..132P, 2012MNRAS.427.3145M, 2013MNRAS.431.1503P, 2013PhRvD..87h3002S, 2013MNRAS.436..449P, 2013MNRAS.433.1529C, 2017MNRAS.472.3959M, 2018JCAP...09..008L, 2018PhR...733....1D}). 

One of the widely used approaches to predicting halo bias parameters on large scales is the peak-background split (PBS) formalism. The underlying idea is that long-wavelength density perturbations modulate the abundance of halos by effectively modifying the background density in which small-scale density fluctuations evolve and eventually collapse to form halos (\cite{1984ApJ...284L...9K, 1986ApJ...304...15B, 1989MNRAS.237.1127C, 1996MNRAS.282..347M, 1998MNRAS.297..692C, 1999MNRAS.308..119S}). In particular, when combined with the excursion set theory, the PBS formalism provides a convenient framework for calculating Lagrangian halo bias parameters at different orders from the FCD. It is worth noting that the PBS formalism yields accurate predictions when implemented exactly (\cite{2018PhR...733....1D}), as demonstrated by \cite{2016JCAP...02..018L}.

The closed-form expression of the first crossing distribution is not known for arbitrary barrier and filter functions. Exceptions include the case of Markovian trajectories  (e.g. \cite{1991ApJ...379..440B, 1998MNRAS.300.1057S}), i.e. trajectories with uncorrelated steps (walks), and completely correlated trajectories (\cite{2012MNRAS.420.1429P}). Assuming Gaussian linear density fluctuations, Markovian trajectories can be generated by applying a ``sharp-k'' smoothing filter, which corresponds to a Top-Hat filter in Fourier space. However, in more realistic scenarios, where filters like real-space Top-Hat, Gaussian and other ones are used, the trajectories become non-Markovian (e.g. \cite{1990MNRAS.243..133P, 1991ApJ...379..440B}). The problem becomes even more complicated when a generic barrier is considered in conjunction with such trajectories to make the analyses consistent with more realistic descriptions of the halo formation process. 

 Given the complexity of non-Markovian stochastic processes with generic barriers, various methods and frameworks have been proposed for approximating the first crossing distribution (e.g. \cite{ 1990MNRAS.243..133P, 2010ApJ...711..907M,  2010ApJ...717..515M, 2011MNRAS.411.2644M,  2011PhRvL.106x1302C, 2011PhRvD..84b3009C, 2012MNRAS.420.1429P, 2012MNRAS.423L.102M, 2013MNRAS.433.3428F,  2014MNRAS.438.2683M, 2018MNRAS.478.5296N, 2024JCAP...10..079V}). In this work, we adopt the Hertz and Stratonovich approximations introduced by \cite{2018MNRAS.478.5296N} to develop a computationally efficient framework for estimating FCDs for generic barriers, which are subsequently used to calculate halo bias parameters within the PBS formalism. We validate the framework by comparing its predictions with those obtained from a numerical evaluation of FCDs, based on large ensembles of excursion set trajectories generated using the Cholesky method (\cite{2018MNRAS.478.5296N}).

This paper is organized as follows: Section \ref{sec:fs_hertz_stra} reviews the Hertz and Stratonovich approximations and uses them to develop an efficient framework for calculating FCDs for generic barriers and smoothing filters. Section \ref{sec:multiplicity_bias_hertz_stra} then discusses how halo bias parameters can be efficiently calculated within the PBS formalism. We subsequently use the framework to compute halo multiplicity functions and biases based on the Hertz and Stratonovich approximations, considering different smoothing filters and barriers. We then compare these predictions with their numerical counterparts. Finally, we conclude this work in Section \ref{sec:conclusions}.

\section{First crossing distribution in Hertz and Stratonovich approximations}
\label{sec:fs_hertz_stra}
The FCD, $f(S)$, can be obtained by calculating the fraction of trajectories that up-cross a barrier for the first time within the interval ($S$, $S+\mathrm{d}S$). However, a useful first step towards computing the FCD would be to first calculate the fraction of trajectories that up-cross the barrier at that interval, regardless of whether they have had previous up-crossings. For this purpose, following the arguments in (\cite{2012MNRAS.423L.102M, 2019PhRvE..99f2101B}), for a moving barrier $B(S)$, one can compute the probability that a trajectory, $\delta(S)$, satisfies the following conditions: $\delta(S) \leq B(S)$ and $\delta(S+ \mathrm{d}S) > B(S + \mathrm{d}S)$. The second condition can be expanded as: $\delta(S) > B(S) + (B^\prime(S) - \eta(S)) \mathrm{d}S$, where $B^\prime  \equiv \mathrm{d}B/\mathrm{d} S$ and $\eta  \equiv \mathrm{d}\delta/\mathrm{d} S$. Assuming that trajectories start from $\delta_0 = 0$ and $\eta_0 = 0$ at $S_0 = 0$, the probability that a trajectory up-crosses the barrier within the interval ($S$, $S+\mathrm{d}S$) can then be written as:
\begin{equation}
	\label{p1v}
	 \int_{B_S + (B^\prime_S - \eta_{_S})\mathrm{d} S}^{B_S} \mathrm{d} \delta_{\tilde{S}}  P(\delta_{\tilde{S}} , \eta_{_{\tilde{S}}} , \tilde{S} )\simeq P(B_S, \eta_{_S}, S)(\eta_{_S} - B^\prime_S) \mathrm{d} S, 
\end{equation}
where $X_S \equiv X(S)$ with $X = (B, B^\prime, \delta, \eta)$. Assuming that linear matter density fluctuations are Gaussian, $\delta_S$ and $\eta_{_S}$ obey Gaussian statistics with $\langle \delta_S \rangle  = \langle \eta_{_S} \rangle = 0$ and $P$ is a joint Gaussian distribution. Now, one can calculate the ``rate'' of up-crossing the barrier at $S$, $n_1(B_S, S)$, which is the probability that a trajectory up-crosses the barrier $B(S)$ per unit $S$, by dividing the result in \eqref{p1v} by $\mathrm{d}S$ and integrating over all slopes that are larger than the slope of the barrier, i.e. $\eta_{_S} > B^\prime_S$. In other words:
\begin{equation}
	\label{n_p1}
	n_1(B_S, S) = \int_{ B^\prime_S}^{\infty} \mathrm{d} \eta_{_S} (\eta_{_S} - B^\prime_S) P(B_S, \eta_{_S}, S).
\end{equation}

The next step in calculating the FCD is to subtract from $n_1$ all contributions corresponding to trajectories that have up-crossed the barrier before $S$ (see e.g. \cite{Verechtchaguina2006, 2014MNRAS.439.3051M, 2018MNRAS.478.5296N, 2019PhRvE..99f2101B}). For this purpose, it is useful to define rate functions for multiple up-crossings, $n_p(B_{S_p}, S_p; \ldots ; B_{S_1}, S_1)$, which correspond to the joint probability that a trajectory up-crosses the barrier at $(S_1, S_1 + \mathrm{d}S_1)$, $\dots$, $(S_p, S_p + \mathrm{d}S_p)$. These rate functions can be expressed by generalizing Eq. \eqref{n_p1}:
\begin{align}
	\label{n_p}
	&n_p(B_{S_p}, S_p; \ldots ; B_{S_1}, S_1) \nonumber\\
	&= \int_{ B^\prime_{S_1}}^{\infty} \mathrm{d} \eta_{_{S_1}} (\eta_{_{S_1}} - B^\prime_{S_1}) \cdots \int_{ B^\prime_{S_p}}^{\infty} \mathrm{d} \eta_{_{S_p}} (\eta_{_{S_p}} - B^\prime_{S_p})\,  \\
	& ~~~~~ \times P(B_{S_p}, \eta_{_{S_p}}, S_p; \ldots; B_{S_1}, \eta_{_{S_1}}, S_1)\,, \nonumber
\end{align}
where $P$ is a multivariate Gaussian distribution. 

To calculate the rate functions, one needs to evaluate auto/cross correlations of $\delta$ and $\eta$ fields. For the $\delta$ correlations we have:
\begin{align}
	\label{cij}
	\langle \delta_i \delta_j \rangle  \equiv  \langle \delta_{S_i} \delta_{S_j}\rangle = &\left \langle \delta(R_i(S_i)) \delta(R_j(S_j))  \right \rangle = \\
	& \int_{0}^{\infty} \mathrm{d}k \frac{k^2 P_{\text{m}}(k)}{2 \pi^2} W(k R_i) W(k R_j) \, , \nonumber
\end{align}
where $P_{\text{m}}(k)$ is the linear matter power spectrum, $R$ denotes the smoothing scale and $W$ is the Fourier transform of the smoothing filter function. The variance $S_i$ is obtained as the special case of this correlation function in the limit $i = j$. From this relation, one can obtain a monotonic mapping between $S$ and $R$, and vice versa, since $S_i = \langle \delta_i \delta_i \rangle = \langle \delta(R_i)^2 \rangle$. The correlations involving the slope variable, $\eta$, can also be evaluated using the following relations:

\begin{align}
	\label{eij1}
	&\langle \eta_i \delta_j \rangle  \equiv  \langle \eta_{_{S_i}} \delta_{S_j}\rangle = \frac{\mathrm{d} R_i}{\mathrm{d} S_i}\left \langle \frac{\mathrm{d} \delta(R_i(S_i))}{\mathrm{d} R_i} \delta(R_j(S_j))  \right \rangle,\\
	&\langle \eta_i \eta_j \rangle  \equiv  \langle \eta_{_{S_i}} \eta_{_{S_j}}\rangle = \frac{\mathrm{d} R_i}{\mathrm{d} S_i} \frac{\mathrm{d} R_j}{\mathrm{d} S_j}\left \langle \frac{\mathrm{d} \delta(R_i(S_i))}{\mathrm{d} R_i} \frac{\mathrm{d} \delta(R_j(S_j))}{\mathrm{d} R_j}  \right \rangle, 
\end{align}
which lead to
\begin{align}
\label{eij2}
 &\langle \eta_i \delta_j \rangle =  \frac{\mathrm{d} R_i}{\mathrm{d} S_i}\left[\int_{0}^{\infty} \mathrm{d}k \frac{k^2 P_{\text{m}}(k)}{2 \pi^2} \frac{\mathrm{d} W(k R_i)}{\mathrm{d} R_i} W(k R_j)\right] , \\ \label{eij3}
&\langle \eta_i \eta_j \rangle = \frac{\mathrm{d} R_i}{\mathrm{d} S_i} \frac{\mathrm{d} R_j}{\mathrm{d} S_j}\left[\int_{0}^{\infty} \mathrm{d}k \frac{k^2 P_{\text{m}}(k)}{2 \pi^2} \frac{\mathrm{d} W(k R_i)}{\mathrm{d} R_i} \frac{\mathrm{d} W(k R_j)}{\mathrm{d} R_j}\right].
\end{align}

Having obtained the rate functions at different orders, corresponding to different numbers of up-crossings, one can obtain the FCD from $n_1$ using the inclusion-exclusion principle (\cite{2019PhRvE..99f2101B}). The final result is an infinite sum whose first term is $n_1$ and the subsequent terms involve integrals of rate functions at different orders. This sum can be expressed as:
\begin{align}
	\label{fs_sum}
	f(S) = \sum_{p = 0}^{\infty} &\frac{(-1)^p}{p!} \int_{0}^{S} \mathrm{d}S_1 \cdots \int_{0}^{S} \mathrm{d}S_p \nonumber  \\ &\times n_{p+1}(B_{S}, S; B_{S_p}, S_p; \ldots ; B_{S_1}, S_1).
\end{align}
It is generally difficult to evaluate this expression exactly. However, the sum can be approximated by treating the contribution of the first few terms exactly and approximating  the contributions of the higher order terms. In this work, we use two approximations of this kind, referred to as  the \emph{Hertz} and \emph{Stratonovich} approximations, which were presented in (\cite{2018MNRAS.478.5296N}). To discuss these further, it is useful to express the FCD in the following way (\cite{Verechtchaguina2006}):
\begin{equation}
	\label{fs_psi}
	f(S) = \psi^\prime(S) e^{-\psi(S)},
\end{equation}
with $\psi^\prime(S)$ denoting a derivative with respect to $S$ and
\begin{equation}
	\label{psi_cumulants}
	\psi(S) \equiv \sum_{p =1}^{\infty} \frac{(-1)^{p+1}}{p!} \int_{0}^{S}\mathrm{d}S_1 \cdots \int_{0}^{S}\mathrm{d}S_p g_p(S_p, \ldots, S_1),
\end{equation}
where $g_p$'s are cumulant functions. For instance, in our case with generic barriers, we have $g_1(S_1) = n_1(B_{S_1}, {S_1})$, $g_2(S_2, S_1) = n_2(B_{S_2}, {S_2}; B_{S_1}, {S_1}) - n_1(B_{S_2}, {S_2}) n_1(B_{S_1}, {S_1})$, etc. 

\subsection{The Hertz approximation}
The idea behind the Hertz approximation (\cite{Hertz1909, 2018MNRAS.478.5296N}) (see also \cite{2014MNRAS.439.3051M}) is that the up-crossings at different points are independent of each other. Under this assumption, the multivariate probability distribution in Eq. \eqref{n_p} can be factorized as $P(B_{S_p}, \eta_{_{S_p}}, S_p; \ldots; B_{S_1}, \eta_{_{S_1}}, S_1) \approx P(B_{S_p}, \eta_{_{S_p}}, S_p)\ldots P(B_{S_1}, \eta_{_{S_1}}, S_1)$. Consequently, this results in the factorization of the rate functions. Therefore, from Eq. \eqref{n_p}, the rate functions in this approximation can be written as:

\begin{align}
	\label{np_factorized}
	&n_p(B_{S_p}, S_p; \ldots ; B_{S_1}, S_1) \approx  \int_{ B^\prime_{S_1}}^{\infty} \mathrm{d} \eta_{_{S_1}} (\eta_{_{S_1}} - B^\prime_{S_1}) P(B_{S_1}, \eta_{_{S_1}}, S_1) 
	\nonumber\\
	&\cdots \int_{ B^\prime_{S_p}}^{\infty} \mathrm{d} \eta_{_{S_p}} (\eta_{_{S_p}} - B^\prime_{S_p})  P(B_{S_p}, \eta_{_{S_p}}, S_p) \nonumber \\
	 &= n_1(B_{S_1}, {S_1}) \ldots n_1(B_{S_p}, {S_p}). 
\end{align}
Now, this result can be inserted into Eq. \eqref{fs_sum} to obtain the FCD in the Hertz approximation:
\begin{align}
	\label{fs_Hz1}
	&f_{\text{Hertz}}(S) = \\
	 &\sum_{p = 0}^{\infty} \frac{(-1)^p}{p!} n_1(B_{S}, {S}) \int_{0}^{S} \mathrm{d}S_1 n_1(B_{S_1}, {S_1})  \cdots \int_{0}^{S} \mathrm{d}S_p n_1(B_{S_p}, {S_p}).\nonumber  
\end{align}
Since the integration limits of each integral in this expression are $0$ and $S$, the integrals are identical and depend only on $S$. In other words:
\begin{align}
	\label{equal_integrals}
	\int_{0}^{S} \mathrm{d}S_i n_1(B_{S_i}, {S_i}) =  \int_{0}^{S} \mathrm{d}S_j n_1(B_{S_j}, {S_j}) = n_1(<S). 
\end{align}
Therefore, Eq. \eqref{fs_Hz1} can be re-expressed as:
\begin{equation}
	\label{fs_Hz}
	f_{\text{Hertz}}(S) = n_1(B_{S}, {S}) e^{- n_1(<S)},
\end{equation}
which, in comparison with Eq. \eqref{fs_psi}, implies that:
\begin{equation}
	\label{psi_hz}
	\psi_{\text{Hertz}} = n_1(<S) = \int_{0}^{S} \mathrm{d}S_1 n_1(B_{S_1}, {S_1}).
\end{equation}
Considering Eq. \eqref{psi_cumulants}, this approximation implies that all cumulant functions vanish except $g_{1}$. 

\subsection{Stratonovich approximation}
The Hertz approximation was the first step toward estimating the sum in Eq. \eqref{fs_sum}. In terms of the cumulant functions, it includes only the contribution of the first cumulant, while approximating higher order terms as zero. The Stratonovich approximation (\cite{Stratonovich1967, 2018MNRAS.478.5296N}) improves upon this estimate by incorporating the contributions of the first two terms exactly and approximating the higher order terms based on these two terms. This approximation can be applied to the cumulant functions as:
\begin{align}
	\label{gp_stra}
	g_p(S_p, \ldots, S_1) &\approx (-1)^{p-1} (p-1)! n_1(B_{S_p}, {S_p}) \ldots n_1(B_{S_1}, {S_1}) \nonumber\\
	& \times \left\{ R(S_1, S_2)R(S_1, S_3)\ldots R(S_1, S_p)  \right\}_{\text{sym}},
\end{align}
where 
\begin{equation}
	\label{Rij}
	 R(S_i, S_j) \equiv  1 - \frac{n_2(B_{S_i}, S_i; B_{S_j}, S_j)}{n_1(B_{S_i}, {S_i})\, n_1(B_{S_j}, {S_j})}.
\end{equation}
This expression can then be inserted into Eq. \eqref{psi_cumulants} and the sum can be evaluated, leading to the Stratonovich approximation for the $\psi$ function (\cite{2018MNRAS.478.5296N}):
\begin{equation}
	\label{psi_Str}
	\psi_{\text{Str}}(S) =  -\int_{0}^{S} \mathrm{d} S_1 n_1(B_{S_1}, S_1) \frac{\ln\left[  1 -  n_1(<S, S_1) \right]}{n_1(<S, S_1)}, 
\end{equation}
where 
\begin{equation}
	\label{n1ssp}
    n_1(<S, S_1) \equiv \int_{0}^{S} \mathrm{d} S_2 R(S_1, S_2) n_1(B_{S_2}, S_2)\,.
\end{equation}
Finally, to find the FCD in Stratonovich approximation, $f_\text{Str}(S)$, one can use Eq. \eqref{fs_psi} with $\psi_{\text{Str}}(S)$. 

\subsection{Calculation of the first crossing distribution}
\label{subsec: n1_n2_fs}
Following the previous parts, the two key functions for calculating the FCD in Hertz and Stratonovich approximations are $ n_1$ and $n_2$ rate functions which can be obtained from Eq. \eqref{n_p}. To find closed-form expressions for them, we make use of the fact that $P$ in Eq. \eqref{n_p} is a multivariate Gaussian distribution. Hereafter, for simplicity in writing, we drop the explicit dependence of Gaussian distributions on $S$ parameters: $P(B_{S_p}, \eta_{_{S_p}}, S_p; \ldots; B_{S_1}, \eta_{_{S_1}}, S_1) = P(B_{S_p}, \eta_{_{S_p}}, \ldots, B_{S_1}, \eta_{_{S_1}})$. 

For $n_1$, one can first write $P(B_{S}, \eta_{_S}) = P(\eta_{_S} | B_{S}) P(B_{S})$. Considering the properties of conditional Gaussian distributions and also using $S = \langle \delta_S^2 \rangle$ and $\langle \delta_S \rangle = \langle\eta_{_S}\rangle = 0$, one obtains
\begin{equation}
	\label{n1p1}
	n_1(B_{S}, S) = \int_{ B^\prime_{S}}^{\infty} \mathrm{d} \eta_{_S} (\eta_{_S} - B^\prime_{S}) P(\eta_{_S} | B_{S}) P(B_{S}),
\end{equation}
where
\begin{equation}
	\label{ps_in_n1}
	P(B_{S}) = \frac{1}{\sqrt{2 \pi S}} e^{-\frac{B_{S}^2}{2 S}}\,, ~~~~~~P(\eta_{_S} | B_{S}) = \frac{1}{\sqrt{2 \pi \sigma_c^2}} e^{-\frac{(\eta_{_S} - \mu_c)^2}{2 \sigma_c^2}}.
\end{equation}
with:
\begin{equation}
	\label{muc_sigc}
    \mu_c \equiv \frac{\langle \delta_S \eta_{_S}\rangle}{S} B_S, ~~~~~~~ \sigma_c^2 \equiv \langle \eta_{_S}^2 \rangle  - \frac{\langle \delta_S \eta_{_S}\rangle^2}{S}.
\end{equation}
The integral in Eq.\eqref{n1p1} can be done analytically. Starting from Eqs \eqref{n1p1} and \eqref{ps_in_n1}, since $B_S^\prime$ is usually a well-behaved function, especially in ranges that are relevant for cosmological analyses, we can apply the change of variable $\eta_{_S} - B_S^\prime = \xi_S$ at each $S$. Therefore, we need to integrate:
\begin{equation}
	\label{n1cv}
	n_1(B_{S}, S) = P(B_{S}) \int_{ 0}^{\infty} \mathrm{d} \xi_{S} \xi_S  \frac{e^{-\frac{(\xi_S + B_S^\prime  - \mu_c)^2}{2 \sigma_c^2}}}{\sqrt{2 \pi \sigma_c^2}} \equiv P(B_S) \times I_S. 
\end{equation}
Now, defining $\tilde{\mu}_c \equiv \mu_c - B_S^\prime$, $I_S$ can be written as:
\begin{align}
	\label{I_s}
	&I_S = \\
	&\frac{\sigma_c^2}{\sqrt{2 \pi \sigma_c^2}} \frac{\partial}{\partial \tilde{\mu}_c} \int_{0} ^{\infty} \mathrm{d} \xi_{S} e^{-\frac{(\xi_S - \tilde{\mu}_c)^2}{2 \sigma_c^2}}  + \frac{ \tilde{\mu}_c}{\sqrt{2 \pi \sigma_c^2}} \int_{0} ^{\infty} \mathrm{d} \xi_{S} e^{-\frac{(\xi_S - \tilde{\mu}_c)^2}{2 \sigma_c^2}}.  \nonumber
\end{align}
One can now apply another change of variables, $ t = (\xi_S - \tilde{\mu}_c)/(\sqrt{2}\sigma_c)$ and re-write the integrals as:
\begin{align}
	\label{I_s2}
	&I_S = \\
	&(\sqrt{2} \sigma_c) \left[\frac{\sigma_c^2}{\sqrt{2 \pi \sigma_c^2}} \frac{\partial}{\partial \tilde{\mu}_c} \int_{-\frac{\tilde{\mu}_c}{\sqrt{2} \sigma_c}} ^{\infty} \mathrm{d} t e^{-t^2}  + \frac{ \tilde{\mu}_c}{\sqrt{2 \pi \sigma_c^2}}\int_{-\frac{\tilde{\mu}_c}{\sqrt{2} \sigma_c}} ^{\infty} \mathrm{d} t e^{-t^2}\right].  \nonumber
\end{align}
The first integral can then be written as:
\begin{equation}
	\label{I_S_1}
	\frac{\sigma_c^2}{\sqrt{2 \pi \sigma_c^2}} \frac{\partial}{\partial \beta} \int_{-\beta} ^{\infty} \mathrm{d} t e^{-t^2} = \frac{\sigma_c}{\sqrt{2 \pi}} e^{-\beta^2},
\end{equation}
where $\beta \equiv \frac{\tilde{\mu}_c}{\sqrt{2} \sigma_c}$. The second integral can also be written as:
\begin{align}
	\label{I_S_2}
	&\frac{ \tilde{\mu}_c (\sqrt{2} \sigma_c)}{\sqrt{2 \pi \sigma_c^2}}\left(\int_{-\frac{\tilde{\mu}_c}{\sqrt{2} \sigma_c}} ^{0} \mathrm{d} t e^{-t^2} + \int_{0} ^{\infty} \mathrm{d} t e^{-t^2}\right)  = \nonumber \\
	&\frac{\tilde{\mu}_c}{2}\left( 1 - \mathrm{erf}\left(-\frac{\tilde{\mu}_c}{\sqrt{2} \sigma_c}\right)\right),
\end{align}
where we have used the definition of the error function. Now, using $\mathrm{erf}(-x) = - \mathrm{erf(x)}$ and combining this result with \eqref{I_S_1} we find the following relation for $I_S$:
\begin{align}
	\label{I_S_f}
	I_S = \frac{\sigma_c}{\sqrt{2 \pi}} e^{-\frac{\tilde{\mu}_c^2}{2 \sigma_c^2}} + \frac{\tilde{\mu}_c}{2}\left( 1 + \mathrm{erf}\left(\frac{\tilde{\mu}_c}{\sqrt{2} \sigma_c}\right)\right),
\end{align}
which, according to Eq. \eqref{n1cv} and the definition of $\tilde{\mu}_c$, leads to the following relation for the $n_1$ rate function:

\begin{align}
	\label{final_n1}
	n_1(B_{S}, S) = &\frac{1}{\sqrt{2 \pi S}} e^{-\frac{B_{S}^2}{2 S}} \left \{  \frac{\sigma_c}{\sqrt{2 \pi} }e^{-\frac{(\mu_c - B_S^\prime)^2}{2 \sigma_c^2}} + \right. \\
	 &\frac{(\mu_c - B_S^\prime)}{2} \left[1 
	\left.+ \mathrm{erf}\left(\frac{\mu_c - B_S^\prime}{\sqrt{2} \sigma_c}\right) \right] \nonumber \right\}.
\end{align}

The next step is to calculate the $n_2$ rate function, which is more involved. However, as will be shown below, the final expressions are compact and straightforward to evaluate. From Eq. \eqref{n_p}, we need to consider the multivariate Gaussian distribution function $P(B_{S_1}, \eta_{_{S_1}}, B_{S_2}, \eta_{_{S_2}})$ which can be expressed as $P( \eta_{_{S_1}},\eta_{_{S_2}} | B_{S_1}, B_{S_2}) P(B_{S_1}, B_{S_2})$ with  $P( \eta_{_{S_1}},\eta_{_{S_2}} | B_{S_1}, B_{S_2})$ being a conditional bivariate Gaussian distribution. This means that we can write the expression for $n_2$ as:
\begin{equation}
	\label{n_2_expression}
	n_2(B_{S_1}, S_1 ; B_{S_2}, S_2) = P(B_{S_1}, B_{S_2})  \times I_{12}(B_{S_1}, S_1 ; B_{S_2}, S_2),
\end{equation}
where
\begin{align}
	\label{I12}
	&I_{12}(B_{S_1}, S_1 ; B_{S_2}, S_2) =  \\
	&\int_{ B^\prime_{S_1}}^{\infty} \mathrm{d} \eta_{_{S_1}} (\eta_{_{S_1}} - B^\prime_{S_1}) \int_{ B^\prime_{S_2}}^{\infty} \mathrm{d} \eta_{_{S_2}} (\eta_{_{S_2}} - B^\prime_{S_2}) P( \eta_{_{S_1}},\eta_{_{S_2}} | B_{S_1}, B_{S_2}). \nonumber
\end{align}
The joint Gaussian distribution in Eq. \eqref{n_2_expression} has a relatively simple expression. It can be obtained from the following relation:

\begin{equation}
	P(B_{S_1}, B_{S_2})
	=  \frac{1}{\sqrt{(2\pi)^2 \det C}}
	\exp\left[
	-\frac{1}{2}
	\left(
	B_{S_1} \, B_{S_2}
	\right)
	C^{-1}
	\begin{pmatrix}
		B_{S_1} \\
		B_{S_2}
	\end{pmatrix}
	\right]
\end{equation}
with
\begin{equation}
		C =  \begin{pmatrix}
		S_1 & \langle \delta_{S_1}\delta_{S_2} \rangle \\
		\langle \delta_{S_2}\delta_{S_1} \rangle  & S_2
	\end{pmatrix}.
\end{equation}
The exponent of the bivariate Gaussian can then be expanded to yield:
\begin{align}
	\label{joint_G}
	&P(B_{S_1}, B_{S_2}) =   \nonumber \\
	&\frac{1}{2\pi\sqrt{S_1 S_2 -  S_{12}^2}} \exp \left[ -\frac{1}{2} \frac{B_{S_2}^2 S_1 + B_{S_1}^2 S_2 - 2 B_{S_1} B_{S_2} S_{12}}{S_1 S_2 -  S_{12}^2} \right] , 
\end{align}
where $S_{12} \equiv \langle \delta_{S_1}\delta_{S_2} \rangle  =  \langle \delta_{S_2}\delta_{S_1} \rangle $. 

Since $ P( \eta_{_{S_1}},\eta_{_{S_2}} | B_{S_1}, B_{S_2})$ in Eq. \eqref{I12} is a conditional bivariate Gaussian distribution, we first briefly review the properties of conditional Gaussian distributions. Consider a vector of  Gaussian random variables, $\vec{x} = (\vec{x}_A, \vec{x}_B)^{\top} $, where $\vec{x}_A \sim \mathcal{N}(\boldsymbol{\mu}_\mathrm{A}, \Sigma_\mathrm{A}) $ and $\vec{x}_B \sim \mathcal{N}(\boldsymbol{\mu}_\mathrm{B}, \Sigma_\mathrm{B}) $ are themselves vectors of Gaussian random variables. Then, one can write $\vec{x} \sim \mathcal{N}(\boldsymbol{\mu}_\mathrm{J}, \Sigma_\mathrm{J})$ with:
\begin{equation}
	\label{J_vars}
	\boldsymbol{\mu}_{\mathrm{J}} = 	\begin{pmatrix}
		\boldsymbol{\mu}_{\mathrm{A}} \\
		\boldsymbol{\mu}_{\mathrm{B}}
	\end{pmatrix} \qquad
	\Sigma_{\mathrm{J}} = \begin{pmatrix}
		\Sigma_{\mathrm{AA}} & \Sigma_{\mathrm{AB}} \\
		\Sigma_{\mathrm{BA}}  & \Sigma_{\mathrm{BB}}
	\end{pmatrix}.
\end{equation}	
For such vectors, the conditional distribution, $P(\vec{x}_A|\vec{x}_B)$, is also a Gaussian,  $P(\vec{x}_A|\vec{x}_B) = \mathcal{N}(\boldsymbol{\mu}, \Sigma) $, with the following mean and variance:
\begin{align}
	\label{conditional_mu_var1}
	&\boldsymbol{\mu} = \boldsymbol{\mu}_A + \Sigma_{\mathrm{AB}} \Sigma_{\mathrm{BB}}^{-1} (\vec{x}_B - \boldsymbol{\mu}_B), \\ \label{conditional_mu_var2}
	& \Sigma = \Sigma_{\mathrm{AA}}  - \Sigma_{\mathrm{AB}} \Sigma_{\mathrm{BB}}^{-1} \Sigma_{\mathrm{BA}}.
\end{align}

For the conditional bivariate Gaussian distribution relevant to our calculations, $ P( \eta_{_{S_1}},\eta_{_{S_2}} | B_{S_1}, B_{S_2})$, we have $\vec{x} = (\eta_{_{S_1}},\eta_{_{S_2}},  \delta_{S_1} = B_{S_1}, \delta_{S_2} = B_{S_2})^{\top}$, $\vec{x}_{\mathrm{A}} = (\eta_{_{S_1}},\eta_{_{S_2}})^{\top}$ and $\vec{x}_{\mathrm{B}} = (\delta_{S_1} = B_{S_1}, \delta_{S_2} = B_{S_2})^{\top}$. Therefore, using Eqs. \eqref{conditional_mu_var1} and \eqref{conditional_mu_var2} for $P(\vec{x}_A|\vec{x}_B)$, one can write:
\begin{equation}
	\label{p12_conditional}
	P( \eta_{_{S_1}},\eta_{_{S_2}} | B_{S_1}, B_{S_2}) = \frac{1}{2 \pi \sqrt{\det \Sigma}} \exp\left[{-\frac{1}{2} (\boldsymbol{\eta} - \boldsymbol{\mu})^{\top}\Sigma^{-1}(\boldsymbol{\eta} - \boldsymbol{\mu})}\right],
\end{equation}
where $\boldsymbol{\eta} = (\eta_{_{S_1}},\eta_{_{S_2}})^{\top}$ and, using $\langle \eta_{_{S_j}} \rangle $ = $\langle \delta_{S_j} \rangle = 0$, 
\begin{equation}
	\label{mu_sigma_conditional}
	\mu = D C^{-1} \begin{pmatrix}
		B_{S_1} \\
		B_{S_2}
	\end{pmatrix}, \qquad \Sigma = E - D C^{-1} D^{\top}
\end{equation}
with 
\begin{align}
	\label{cde}
	&C =  \begin{pmatrix}
		S_1 & \langle \delta_{S_1}\delta_{S_2} \rangle \\
		\langle \delta_{S_2}\delta_{S_1} \rangle  & S_2
	\end{pmatrix} \equiv  \begin{pmatrix}
		S_1 & S_{12}\\
		S_{21} & S_2
	\end{pmatrix}, \qquad \\
	&D =  \begin{pmatrix}
		\langle \eta_{_{S_1}} \delta_{S_1} \rangle & \langle \eta_{_{S_1}}\delta_{S_2}\rangle \\
		\langle \eta_{_{S_2}} \delta_{S_1} \rangle  & \langle \eta_{_{S_2}} \delta_{S_2} \rangle
	\end{pmatrix}  \equiv \begin{pmatrix}
		\mathrm{ed}_{11} & \mathrm{ed}_{12} \\
		\mathrm{ed}_{21}  & \mathrm{ed}_{22}
	\end{pmatrix},   \\
	& E = \begin{pmatrix}
		\langle \eta_{_{S_1}} \eta_{_{S_1}} \rangle & \langle \eta_{_{S_1}} \eta_{_{S_2}} \rangle \\
		\langle \eta_{_{S_2}}\eta_{_{S_1}} \rangle  & \langle \eta_{_{S_2}} \eta_{_{S_2}} \rangle
	\end{pmatrix} \equiv \begin{pmatrix}
		\mathrm{e}_{1}& \mathrm{e}_{12} \\
		\mathrm{e}_{21} & \mathrm{e}_{2}
	\end{pmatrix}. 
\end{align}
In these expressions, we also have $S_{12} = S_{21}$ and $\mathrm{e}_{12} = \mathrm{e}_{21}$. From these matrices, one can obtain the elements of $\boldsymbol{\mu}$ and $\Sigma$ introduced in \eqref{mu_sigma_conditional}.  The final results are:
\begin{equation}
	\label{mu1_final}
	\mu_1 = \frac{\mathrm{ed}_{11}  (B_{S_2} S_{12} - B_{S_1} S_2) + \mathrm{ed}_{12} (B_{S_1} S_{12} - B_{S_2} S_1)}{S_{12}^2 - S_1 S_2},
\end{equation}
\begin{equation}
	\label{mu2_final}
	\mu_2 = \frac{\mathrm{ed}_{21}  (B_{S_2} S_{12} - B_{S_1} S_2) + \mathrm{ed}_{22} (B_{S_1} S_{12} - B_{S_2} S_1)}{S_{12}^2 - S_1 S_2},
\end{equation}
\begin{equation}
	\label{sigma_11_final}
	\Sigma_{11} =  \frac{\mathrm{ed}_{12}^2 S_1 - 2 \mathrm{ed}_{11} \mathrm{ed}_{12} S_{12} + \mathrm{e}_{1}^2 S_{12}^2 +  \mathrm{ed}_{11}^2 S_2  -  \mathrm{e}_{1}^2 S_{1} S_{2} }{S_{12}^2 - S_1 S_2}, 
\end{equation}
\begin{equation}
	\label{sigma_22_final}
	\Sigma_{22} = \frac{\mathrm{ed}_{22}^2 S_1 - 2 \mathrm{ed}_{21} \mathrm{ed}_{22} S_{12} + \mathrm{e}_{2}^2 S_{12}^2 +  \mathrm{ed}_{21}^2 S_2  -  \mathrm{e}_{2}^2 S_{1} S_{2} }{S_{12}^2 - S_1 S_2},
\end{equation}
\begin{align}
	\label{sigma_12_final}
	\Sigma_{12} = \Sigma_{21} = &\frac{1}{S_{12}^2 - S_1 S_2}\bigg(\mathrm{ed}_{12}(\mathrm{ed}_{22} S_1  - \mathrm{ed}_{21} S_{12}) +  \\ 
	&\mathrm{ed}_{11}(\mathrm{ed}_{21} S_2 - \mathrm{ed}_{22} S_{12}) + 
	\mathrm{e}_{12}(S_{12}^2 - S_1 S_2) \bigg). \nonumber
\end{align}
These equations provide everything needed to fully determine the conditional bivariate Gaussian distribution in Eq. \eqref{p12_conditional}. 

The next step is obtaining a closed-form expression for the $I_{12}$ function. From Eqs \eqref{I12} and \eqref{p12_conditional}, one can re-write the $I_{12}$ integral as:
\begin{align}
	\label{I12_rwr}
	&I_{12} = \int_{ B^\prime_{S_1}}^{\infty} \mathrm{d} \eta_{_{S_1}} (\eta_{_{S_1}} - B^\prime_{S_1}) \int_{ B^\prime_{S_2}}^{\infty} \mathrm{d} \eta_{_{S_2}} (\eta_{_{S_2}} - B^\prime_{S_2}) \\
	&\times \left(\frac{1}{2 \pi \sqrt{\Sigma_{11} \Sigma_{22} - \Sigma_{12}^2}} \exp{\left[\frac{\Sigma_{11} \Delta \eta_2^2 + \Sigma_{22} \Delta \eta_1^2 - 2 \Sigma_{12} \Delta \eta_1 \Delta \eta_2 }{2\left(\Sigma_{12}^2 - \Sigma_{11} \Sigma_{22}\right)}\right]} \right), \nonumber
\end{align}
where $\Delta \eta_j \equiv \eta_{_{S_j}} - \mu_j$.  By applying the following change of variables:
\begin{equation}
	u = \frac{\Delta \eta_1}{\sqrt{\Sigma_{11}}} \qquad v = \frac{\Delta \eta_2}{\sqrt{\Sigma_{22}}},
\end{equation}
and defining:
\begin{equation}
	\sigma_1 \equiv  \sqrt{\Sigma_{11}} \qquad \sigma_2 \equiv  \sqrt{\Sigma_{22}} \qquad \rho \equiv  \frac{\Sigma_{12}}{\sigma_1 \sigma_2},
\end{equation}
after some algebra, the integral takes the following form:
\begin{align}
	\label{I12_rwr2}
	&I_{12} = \\
	&\int_{(\frac{B^\prime_{S_1} - \mu_1}{\sigma_1})}^{\infty} \mathrm{d} u (u \sigma_1 + \mu_1 - B^\prime_{S_1}) \int_{(\frac{B^\prime_{S_2} - \mu_2}{\sigma_2})}^{\infty} \mathrm{d} v (v \sigma_2 + \mu_2 - B^\prime_{S_2}) \nonumber \\
	&\times \left(\frac{1}{2 \pi \sqrt{1 - \rho^2}} \exp{\left[-\frac{u^2 + v^2 - 2 u v \rho}{2\left(1 - \rho^2\right)}\right]} \right). \nonumber
\end{align}
Now, we define
\begin{equation}
	\label{a1a2}
	a_1 \equiv \frac{B^\prime_{S_1} - \mu_1}{\sigma_1}, \qquad a_2  \equiv \frac{B^\prime_{S_2} - \mu_2}{\sigma_2},
\end{equation}	
and the integral is further simplified as
\begin{align}
	\label{I12_rwr2}
	I_{12} = &\sigma_1 \sigma_2 \int_{a_1}^{\infty} \mathrm{d} u (u - a_1) \int_{a_2}^{\infty} \mathrm{d} v (v - a_2)  \\
	&\times \left(\frac{1}{2 \pi \sqrt{1 - \rho^2}} \exp{\left[-\frac{u^2 + v^2 - 2 u v \rho}{2\left(1 - \rho^2\right)}\right]} \right) \nonumber  \\ 
	&\equiv \sigma_1 \sigma_2 J_{12}(a_1, a_2; \rho). \nonumber
\end{align}
Note that the expression inside the parentheses is the probability density function (PDF) of a standard bivariate Gaussian distribution, which is a well-known function in statistics. Referring to this expression as $\phi_2(u, v; \rho)$, the integrand of $J_{12}(a_1, a_2; \rho)$ can be expanded:
\begin{align}
	\label{J12}
	&J_{12}(a_1, a_2; \rho) = \int_{a_1}^{\infty} \mathrm{d} u  \int_{a_2}^{\infty} \mathrm{d} v  (u v) \phi_2(u, v; \rho) \\
	&- a_2 \int_{a_1}^{\infty} \mathrm{d} u  \int_{a_2}^{\infty} \mathrm{d} v  \,u \phi_2(u, v; \rho) - a_1\int_{a_1}^{\infty} \mathrm{d} u  \int_{a_2}^{\infty} \mathrm{d} v  \,v\phi_2(u, v; \rho) \nonumber \\
	& + a_1 a_2 \int_{a_1}^{\infty} \mathrm{d} u  \int_{a_2}^{\infty} \mathrm{d} v  \phi_2(u, v; \rho). \nonumber
\end{align}
Here, the \textbf{last integral} gives the probability $P_2(u > a_1, v > a_2)$. Similarly, the first three integrals correspond to different ``moments'' of the standard bivariate Gaussian PDF over the desired region. Such moments have already been calculated in the mathematical literature. Here we use the results of (\cite{Rosenbaum1961}), which, after some algebra, lead to the following expression for the first three terms:
\begin{align}
	\label{J12_f3}
	& -a_1 \phi(a_2)\, \Phi(\frac{\rho a_2 - a_1}{\sqrt{1 - \rho^2}})  - a_2 \phi(a_1)\, \Phi(\frac{\rho a_1 - a_2}{\sqrt{1 - \rho^2}})  \\
	&+ \frac{\sqrt{1 - \rho^2}}{\sqrt{2 \pi}} \phi \left(\frac{\sqrt{a_1^2 + a_2^2 - 2 a_1 a_2 \rho}}{\sqrt{1 - \rho^2}}\right) + \nonumber \\
	&\rho \int_{a_1}^{\infty} \mathrm{d} u  \int_{a_2}^{\infty} \mathrm{d} v  \phi_2(u, v; \rho), \nonumber
\end{align}
where $\phi$ and $\Phi$ are, respectively, the PDF and cumulative distribution function (CDF) of a one-dimensional standard Gaussian distribution. The probability integral $P_2(u > a_1, v > a_2)$ appears here again in the last term. This can be expanded as:
\begin{align}
	&P_2(u > a_1,v > a_2)\\ 
	&= 1 - P_2(u\leq a_1) - P_2(v \leq a_2)
	+ P_2(u\leq a_1,v\leq a_2). \nonumber
\end{align}
Therefore, for the integral we have:
\begin{align}
	&\int_{a_1}^{\infty} \mathrm{d} u  \int_{a_2}^{\infty} \mathrm{d} v  \phi_2(u, v; \rho) =  P_2(u > a_1,v > a_2)\\
	&= 1 - \Phi(a_1) - \Phi(a_2)  + \Phi_2(a_1, a_2; \rho), \nonumber
\end{align}
where $\Phi_2(a_1, a_2; \rho)$ is the CDF of a standard bivariate Gaussian distribution with correlation coefficient $\rho$. From this expression and \eqref{J12_f3} for the first three terms of $J_{12}$, one can obtain the final expression as:
\begin{align}
	\label{J12_final}
	&J_{12}(a_1, a_2; \rho) = \\
	& -a_1 \phi(a_2)\, \Phi(\frac{\rho a_2 - a_1}{\sqrt{1 - \rho^2}})  - a_2 \phi(a_1)\, \Phi(\frac{\rho a_1 - a_2}{\sqrt{1 - \rho^2}})  \nonumber \\
	&+ \frac{\sqrt{1 - \rho^2}}{\sqrt{2 \pi}} \phi \left(\frac{\sqrt{a_1^2 + a_2^2 - 2 a_1 a_2 \rho}}{\sqrt{1 - \rho^2}}\right) \nonumber \\
	&+ (\rho + a_1 a_2) \left(1 - \Phi(a_1) - \Phi(a_2) + \Phi_2(a_1, a_2; \rho)\right). \nonumber
\end{align}
Finally, since we had defined $I_{12} \equiv \sigma_1 \sigma_2 J_{12}(a_1, a_2; \rho)$, this results in:

 \begin{align}
 	\label{I12_final}
 	&I_{12}(B_{S_1}, S_1 ; B_{S_2}, S_2) = \\
 	&\sigma_1 \sigma_2 \Bigg\{ -a_1 \phi(a_2)\, \Phi(\frac{\rho a_2 - a_1}{\sqrt{1 - \rho^2}})  - a_2 \phi(a_1)\, \Phi(\frac{\rho a_1 - a_2}{\sqrt{1 - \rho^2}})  \nonumber \\
 	&~~~~~~~~~+ \frac{\sqrt{1 - \rho^2}}{\sqrt{2 \pi}} \phi \left(\frac{\sqrt{a_1^2 + a_2^2 - 2 a_1 a_2 \rho}}{\sqrt{1 - \rho^2}}\right) \nonumber \\
 	&~~~~~~~~~+ (\rho + a_1 a_2) \left(1 - \Phi(a_1) - \Phi(a_2) + \Phi_2(a_1, a_2; \rho)\right)    \Bigg\}, \nonumber
 \end{align}
 based on which the rate function $n_2$ was presented in Eq. \eqref{n_2_expression}. We recall that $\phi$ and $\Phi$ denote, respectively, the PDF and CDF of a one-dimensional standard Gaussian distribution and $\Phi_2(a_1, a_2; \rho)$ denotes the CDF of a standard bivariate Gaussian distribution. All these functions are well-known and have been implemented in most standard computational packages. The five parameters in this expression are:
\begin{align}
	\label{I12_params}
	&\sigma_1 \equiv \sqrt{\Sigma_{11}},\,\, \,\,\,\, \sigma_2 \equiv \sqrt{\Sigma_{22}},\,\, \,\,\,\, \rho \equiv \frac{\Sigma_{12}}{\sigma_1 \sigma_2},\,\, \,\,\,\,  \\
	 & a_1 \equiv \frac{B^\prime_{S_1} - \mu_1}{\sigma_1},\,\, \,\,\,\,  a_2 \equiv \frac{B^\prime_{S_2} - \mu_2}{\sigma_2},
\end{align}
where $\Sigma_{11}$, $\Sigma_{22}$, $\Sigma_{12}$, $\mu_1$ and $\mu_2$ can be found in Eqs \eqref{mu1_final} - \eqref{sigma_12_final}. These are functions that, in general, depend on $\langle \delta_{S_i} \delta_{S_j}\rangle$, $\langle \eta_{S_i} \eta_{S_j}\rangle$,$\langle \eta_{S_i} \delta_{S_j}\rangle$ and $B_{S_{i}}$ where each of $i$ and $j$ can be $1, 2$. 

\subsection{Summary of the calculation steps}
\label{subsection:summary_fs}
We are now able to calculate the FCD based on the Hertz and Stratonovich approximations for any given smoothing filter, barrier function, and set of cosmological parameters. The first step is to calculate the correlation functions from Eqs \eqref{cij}, \eqref{eij2} and \eqref{eij3} for a given filter function. The next step is to calculate the $n_1$ rate function using Eq. \eqref{final_n1}. Next, the $n_2$ rate function can be calculated using Eq. \eqref{n_2_expression}, where the expressions appearing in this equation are given by Eqs \eqref{joint_G} and \eqref{I12_final}. Given the forms of these expressions, all steps are computationally straightforward and can be performed efficiently. Finally, after calculating the $n_1$ and $n_2$ rates, one can use Eqs \eqref{psi_hz} and \eqref{psi_Str} to obtain the $\psi$ functions in Hertz and Stratonovich approximations, respectively, and calculate the corresponding FCDs from Eq. \eqref{fs_psi}. All these steps are also numerically straightforward and computationally efficient.

\section{Halo multiplicity function and bias approximations}
\label{sec:multiplicity_bias_hertz_stra}

After calculating the FCD for a given barrier and filter function following the steps outlined in the previous subsection, one can readily determine the corresponding halo multiplicity function --- a quantity widely used to calculate halo abundances. Halo multiplicity function is usually expressed as $\nu f(\nu)$, where $\nu \equiv \delta_c / \sqrt{S}$ denotes the peak height parameter and $\delta_c$ is the critical overdensity for collapse (halo formation threshold). The function $f(\nu)$ can be calculated from the FCD using the relation $f(S) dS = -f(\nu) d \nu$, which implies that $f(\nu)$ is a normalized distribution with an interpretation similar to that of $f(S)$, but for $\nu$ intervals. Considering the relation between $\nu$ and $S$, one obtains:
\begin{equation}
	\label{fs_to_fnu}
	f(\nu) = 2 \frac{\delta_c^2}{\nu^3}f(S(\nu)),
\end{equation}
where $S(\nu) = \delta^2_c/\nu^2$. 

One motivation for using $f(\nu)$ instead of $f(S)$ is the fact that early models predicted the multiplicity function to be universal, i.e. absorbing the cosmology and redshift dependence into the peak height parameter $\nu$ (e.g. \cite{1974ApJ...187..425P, 1999MNRAS.308..119S, 2001MNRAS.321..372J}). However, later studies based on higher-resolution N-body simulations demonstrated that universality is not exact (e.g. \cite{2006ApJ...646..881W, 2007MNRAS.374....2R, 2008ApJ...688..709T, 2010MNRAS.403.1353C}). Furthermore, EST-based predictions for $f(\nu)$ for arbitrary choices of smoothing filter and barrier functions are not guaranteed to be universal either. Nonetheless, using $f(\nu)$ is still useful as it is near-universal in many cases and, more generally, can provide an efficient framework for modeling departures from universality (e.g. \cite{2016MNRAS.456.2486D}). The relation between the halo multiplicity function and the halo mass function, the comoving number density of a halos per unit $\log M$ interval, can be written as:

\begin{equation}
\label{mass_function}
\frac{\mathrm{d} n }{\mathrm{d} \log M} = \frac{\bar{\rho}}{M} f(S) \left|\frac{\mathrm{d} S}{\mathrm{d} \log M}\right|= \frac{\bar{\rho}}{M} \left(\nu f(\nu)\right) \frac{\mathrm{d} \ln \nu}{\mathrm{d} \log M},	
\end{equation}
where $\bar{\rho}$ denotes the average comoving matter density. 

In addition to the multiplicity function and halo mass function, one can also estimate the halo bias from the FCD for a given collapse barrier, $B(S, \delta_c)$\footnote{Here we consider $\delta_c$ to be a general halo formation threshold.}, and a chosen smoothing filter. In general, the parameters of the barrier, including those determining its shape, can be calibrated against N-body simulations (e.g. \cite{2001MNRAS.323....1S, 2024JCAP...10..079V}). Here, we focus on the calculation of halo bias using the PBS formalism (see the references in Section \ref{sec:intro}). Given that halo bias can be interpreted as the response of halo abundance to a long-wavelength background density fluctuation, and considering the abundance formula in Eq. \eqref{mass_function}, one can calculate the Lagrangian bias parameters within the EST framework by studying the conditional function $f(S| S_\ell, \delta_\ell)$, which is the FCD for trajectories conditioned on crossing the point ($S_\ell, \delta_\ell$), where $\delta_\ell$ is the long-wavelength background density fluctuation and $S_\ell$ is its variance. In fact, the Lagrangian halo density contrast can be written as
\begin{equation}
	\label{lagrangian_bias}
	1 + \delta_h^{(\mathrm{L})}  = \frac{f(S| S_\ell, \delta_\ell)}{f(S| 0, 0)},
\end{equation}
where $f(S| 0, 0) = f(S)$ is the unconditional FCD. At large scales where $S_\ell \ll 1$, one can obtain local Lagrangian bias parameters by expanding the conditional FCD in terms of $\delta_\ell$. In other words:
\begin{equation}
	\label{lagrangian_bias2}
	 \delta_h^{(\mathrm{L})}  =   \sum_{n = 1}^{\infty} 
	\frac{b^{(\mathrm{L})}_n}{n !} \delta_{\ell}^n  =  \frac{1}{f(S)} \sum_{n = 1}^{\infty} \frac{1}{n !} \frac{\partial^n  f(S| 0, \delta_\ell)}{\partial \delta^{n}_{\ell}} \bigg |_{\delta_\ell = 0} \delta^{n}_\ell ,
\end{equation}
which leads to
\begin{equation}
	\label{lagrangian_bias3}
	b^{(\mathrm{L})}_n = \frac{1}{f(S)} \frac{\partial^n  f(S| 0, \delta_\ell)}{\partial \delta^{n}_{\ell}} \bigg |_{\delta_\ell = 0}.
\end{equation}
These Lagrangian bias parameters can then be transformed into Eulerian bias parameters at different orders using appropriate transformation relations (e.g. \cite{1998MNRAS.297..692C, 2011PhRvD..83h3518M, 2018JCAP...07..029A}). 
\begin{figure*}
	\centering
	\includegraphics[width=\textwidth]{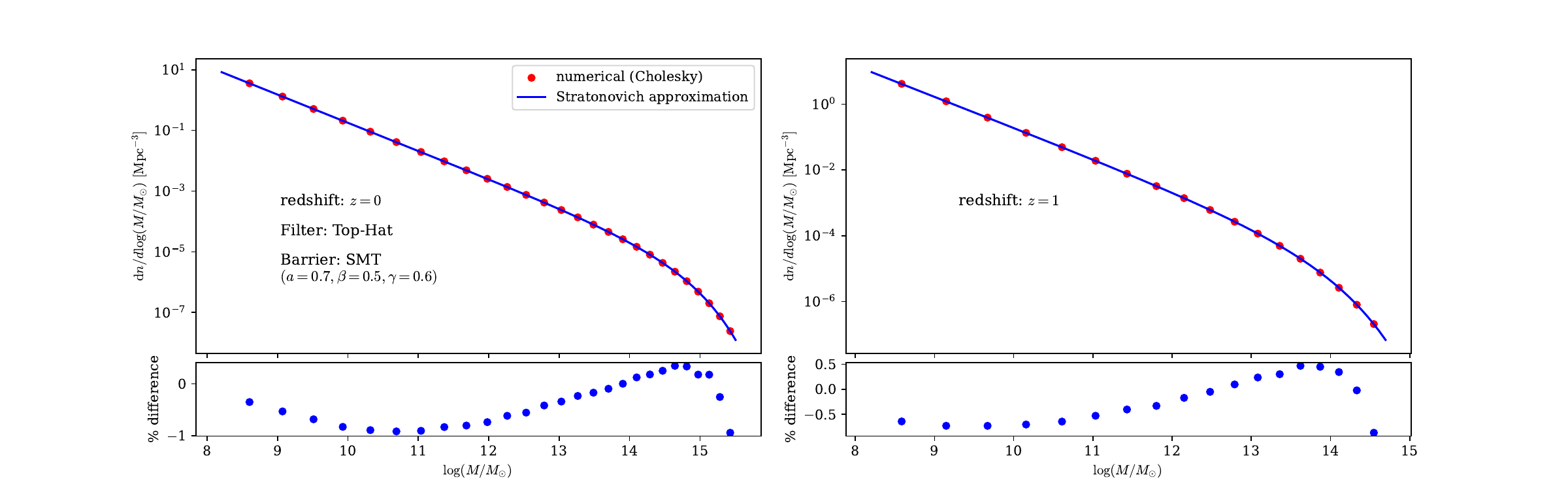}
	\caption{halo mass functions calculated from Eq. \eqref{mass_function}, based on first crossing distributions calculated from the Stratonovich approximation (blue solid lines) and numerically generated excursion set trajectories using Cholesky method (red points), at two redshifts. The botton panels show the percent difference between the Stratonovich and numerical results. The range of halo masses are chosen such that the statistical errors of the numerical values are sub-percent.}
	\label{fig:fscmpr}
\end{figure*}
\begin{figure*}
	\centering
	\includegraphics[width = \textwidth]{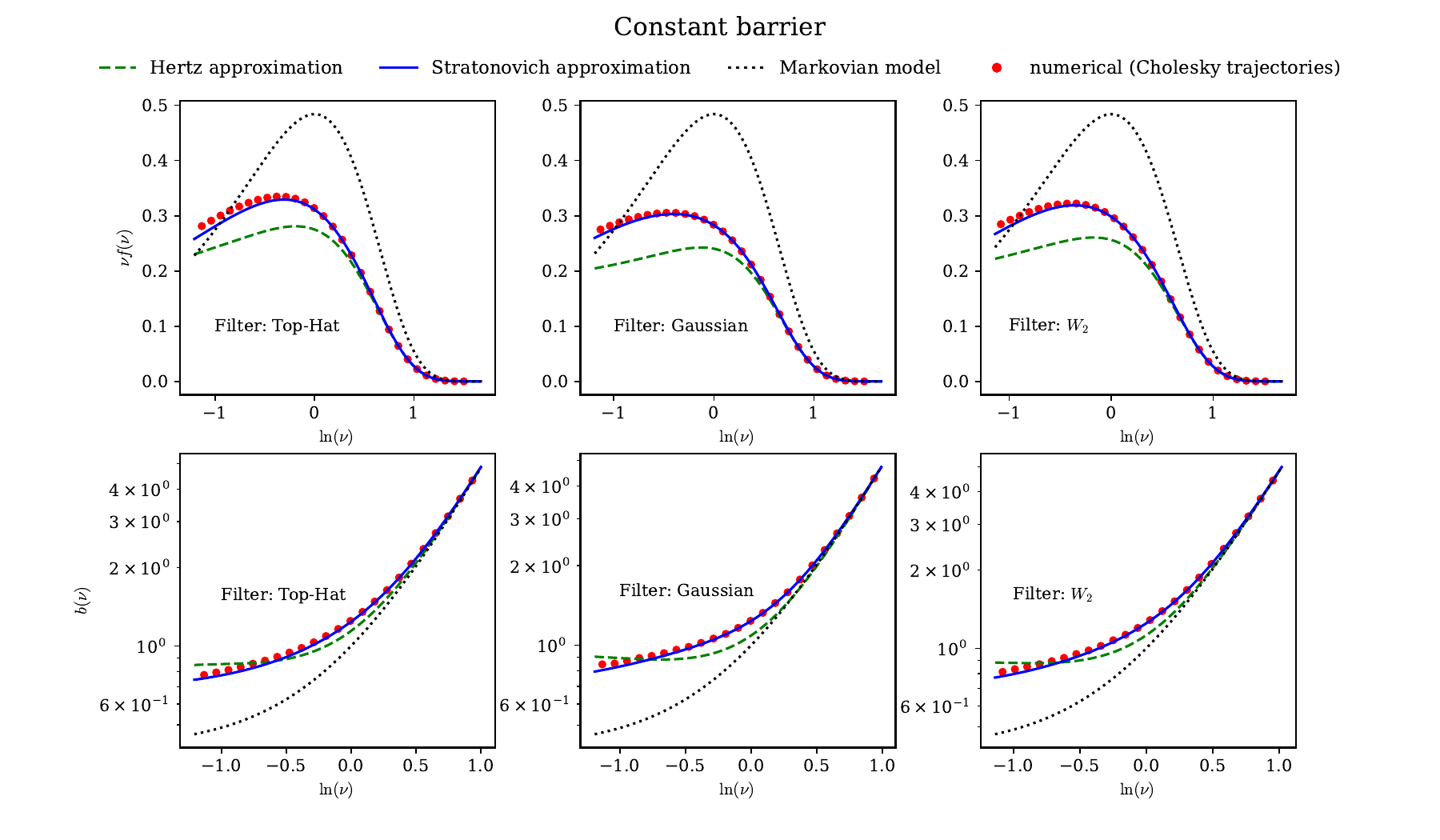}
	\caption{Halo multiplicity functions and biases obtained from different approaches for different smoothing filters, assuming a constant barrier $B = \delta_c = 1.686$. The predictions are based on the Hertz (green dashed lines) and Stratonovich (blue solid lines) approximations, compared with their numerical counterparts based on numerically generated excursion set trajectories using the Cholesky method (red points) and also Markovian multiplicity function and bias results based on Eqs \eqref{nufnupsh} and \eqref{bMmw} (black dotted line).}
	\label{fig:Bc}
\end{figure*}
\begin{figure*}
	\centering
	\includegraphics[width = \textwidth]{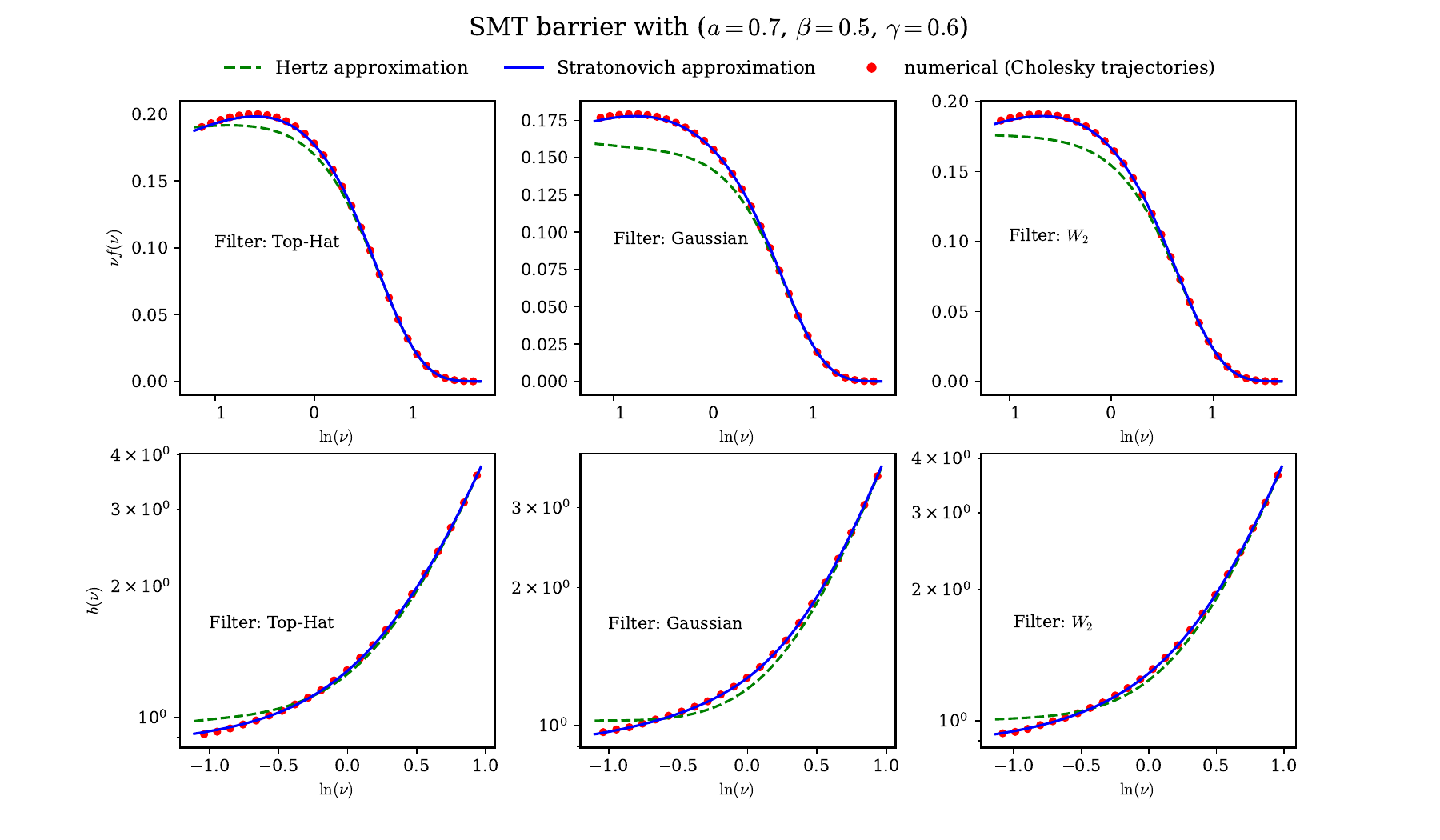}
	\caption{Halo multiplicity functions and biases obtained from different approaches for different smoothing filters, assuming an SMT moving barrier with the parameters specified in the title.}
	\label{fig:SMT756}
\end{figure*}

For uncorrelated (Markovian) steps, the conditional FCD in Eq. \eqref{lagrangian_bias} can be replaced with an unconditional FCD with appropriate shifts in $S$ and the barrier (e.g. \cite{2012MNRAS.419..132P}). In fact, for Markovian excursion set trajectories that first up-cross a barrier of the general form, $B(S) = B_c + B_{s}(S)$,  the conditional FCD, $f(S| S_\ell, \delta_\ell)$, can be replaced with an unconditional FCD, $f(\tilde{S})$ which is obtained from the FCD of trajectories for a barrier modified as $B(\tilde{S} + S_{\ell}) - \delta_{\ell}$ with $\tilde{S} = S - S_{\ell}$ (here we assume that no trajectories, or only a negligible number of trajectories, have crossed the barrier at $s < S_\ell$). At the large scale limit ($S_\ell \ll 1$) for $S_\ell \ll S$, $f(\tilde{S})$ can be simply approximated as $f(S)$ for a barrier shifted as $B(S) \to B(S) - \delta_{\ell}$. This implies that in the Markovian case, the local bias parameters at large scales can be obtained from an equation similar to Eq. \eqref{lagrangian_bias3}, but with the conditional FCD replaced by the unconditional one described above and the derivatives taken with respect to $- B_c$ instead of $\delta_\ell$. In other words, for uncorrelated trajectories, the large scale Lagrangian local bias parameters can be attributed to the response of the $f(S)$ function to variations of the $B_c$. 

The situation, however, becomes different for correlated (non-Markovian) trajectories. Following (\cite{2012MNRAS.419..132P, 2012MNRAS.427.3145M, 2018PhR...733....1D}), for non-Markovian trajectories, the large scale ($S_\ell \ll 1$) Lagrangian bias parameters can be well approximated  as:
	\begin{equation}
		\label{lagrangian_bias4}
		b^{(\mathrm{L})}_n = \left( \frac{S_{\times}}{S_\ell}\right)^n b^\mathrm{(L)}_{n0}.
	\end{equation}
	Here, $S_{\times} \equiv \langle \delta(S) \delta_{\ell}(S_{\ell})\rangle$ and also $S_{\ell} = \langle \delta_{\ell}^2\rangle$ can be calculated from Eq. \eqref{cij} for  Lagrangian radii $R$ (corresponding to $S$) and $R_\ell \gg R$ (corresponding to $S_\ell$). The ratio $S_{\times} / S_{\ell}$ is an $\mathcal O(1)$ quantity which becomes equal to  $1$ for Markovian trajectories (sharp-k filter). The $b^\mathrm{(L)}_{n0}$ parameters, which are distinguished from other bias parameters that are multiplied by scale-dependent quantities that become negligible at $S_\ell \ll 1$ limit, can be obtained from 
	\begin{equation}
		\label{bn0}
		b^{\mathrm{(L)}}_{n0} = \frac{(-1)^n}{f(S)} \frac{\partial^n  f(S)}{\partial B^{n}_c}.
	\end{equation}
This relation is motivated by the first order bias relation for non-Markovian trajectories in the presence of  a moving barrier derived in \cite{2012MNRAS.419..132P}, as well as by the bias relations at different orders for non-Markovian trajectories in the presence of a constant barrier derived in \cite{2012MNRAS.427.3145M, 2018PhR...733....1D}. Eqs \eqref{lagrangian_bias4} and \eqref{bn0} recover the correct limits both for a constant barrier and when non-Markovian contributions become negligible. 

Finally it is worth noting that these biases are real space bias parameters. In general, scale dependent contributions to bias parameters give rise to $k$-dependent terms in their Fourier space relations. The large scale limit of the halo bias in Fourier space can then be recovered in the limit $k \ll 1\, \mathrm{Mpc^{-1}}$. Considering the full $k$-dependent expression for the halo bias in Fourier space in (\cite{2012MNRAS.427.3145M}), at linear order one can write: $\delta^\mathrm{(L)}_{h}(k) \simeq b^\mathrm{(L)}_{10} W(k R) \delta^{\mathrm{(L)}}_{\mathrm{lin}}(k)$  in the limit $k \ll 1\, \mathrm{Mpc^{-1}}$. This implies that $b^\mathrm{(L)}_{10} $ can be interpreted as the Fourier space Lagrangian linear bias parameter on large scales.

The framework provided in this work for calculation of the FCD (summarized in subsection \ref{subsection:summary_fs}) can be used to efficiently calculate the $b^{\mathrm{(L)}}_{n0}$ parameters and obtain the corresponding Lagrangian bias parameters from Eq. \eqref{lagrangian_bias4}. The procedure for obtaining the $b^{\mathrm{(L)}}_{n0}$ parameters from Eq. \eqref{bn0} can be described as follows: 

\begin{itemize}          
	\item For a given barrier of the form $B(S) = B_c + B_{s}(S)$ and a given  filter function, one can first calculate the unconditional FCDs for a range of values of $B_c$ around the desired fiducial value. This provides, at each $S$, a range of values of $f(S)$ corresponding to different values of $B_c$ at and around the fiducial value.
	
	\item After calculating the corresponding $f(S)$ functions, one can evaluate their derivatives with respect to $B_c$ at any $S$ using straightforward numerical procedures, and obtain the corresponding $b^{\mathrm{(L)}}_{n0}$ values from Eq. \eqref{bn0}. 
	
	\item Then, the results can be inserted into Eq. \eqref{lagrangian_bias4} to obtain the corresponding bias parameters. 
\end{itemize}

This procedure can be used to obtain the large scale bias parameters for general chcices of the barrier and filter functions. However, in some cases, these parameters can be calculated from more universal expressions. Following the discussions and results in (\cite{2018PhR...733....1D}) on the derivation of the Lagrangian bias parameters based on the exact implementation of the PBS formalism through the separate-universe approach, under certain conditions and assumptions, one can find:  
\begin{equation}
	\label{nu_PBS}
	b^{\mathrm{(L)}}_{n} = \frac{(- \nu)^n}{(\delta_c)^{n}} \frac{1}{\nu f(\nu)} \frac{\mathrm{d}^n \left(\nu f(\nu)\right)}{d \nu^n}.
\end{equation}
One of the key assumptions underlying the derivation of this relation is the universality of the multiplicity function, $\nu f(\nu)$. That is, all the cosmological dependence is assumed to be encapsulated in the peak height parameter $\nu$, with no additional dependence on the critical density, $\delta_c$. In addition, although this relation is derived for a constant barrier, it can also be extended to a moving barrier, $B(S)$, provided that the barrier is assumed to remain unchanged in the presence of a long-wavelength fluctuation (\cite{2018PhR...733....1D}). It should, however, be noted that this is, in general, an approximation since physical effects governing the behavior of a moving barrier (e.g. tidal effects) can potentially be influenced by a background fluctuation. 
 \begin{figure*}
 	\centering
 	\includegraphics[width=\textwidth]{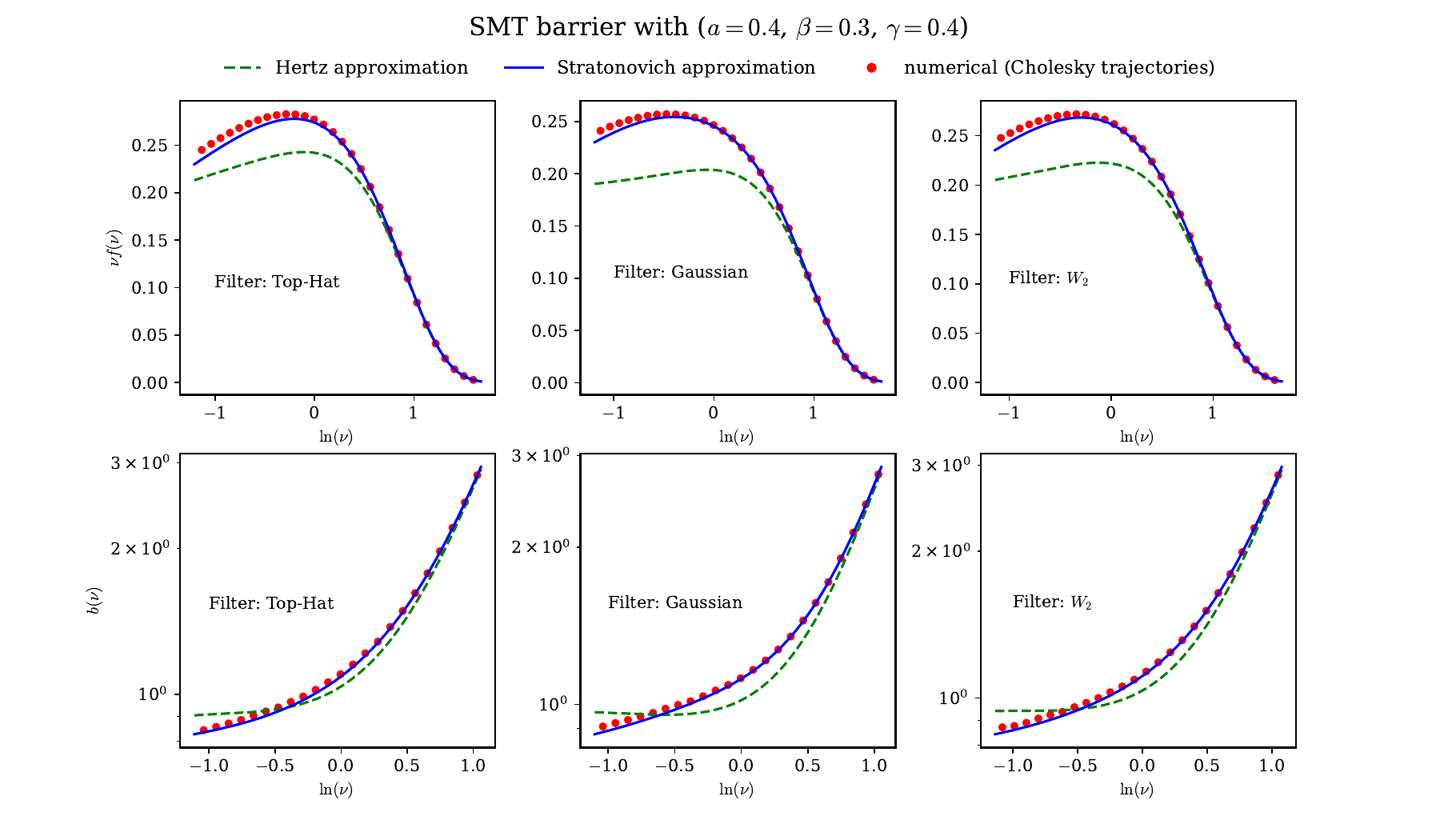}
 	\includegraphics[width=\textwidth]{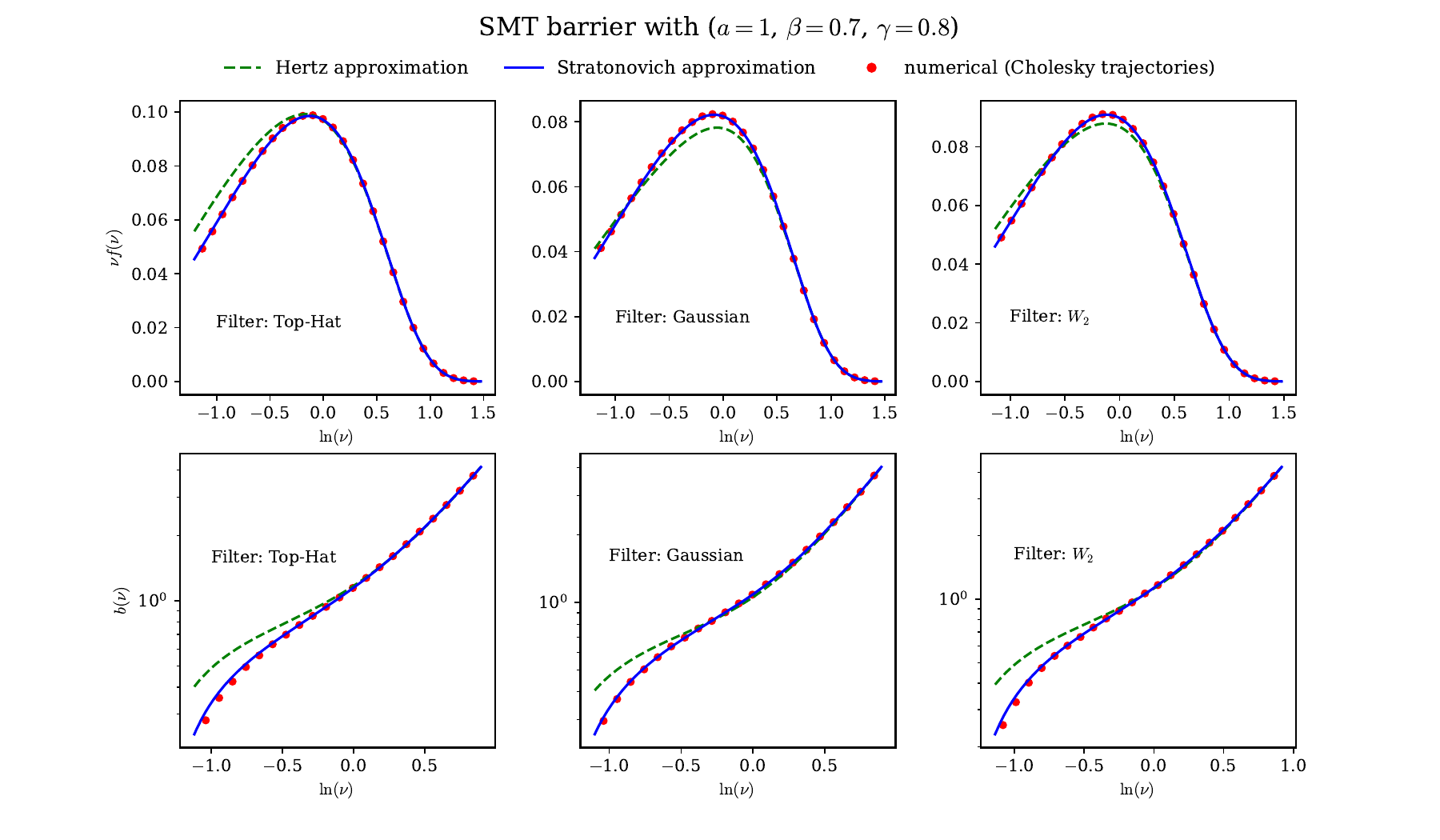}
 	\caption{Similar to Fig. \ref{fig:SMT756}, but for different values of the SMT barrier parameters.}
 	\label{fig:SM434_178}
 \end{figure*}
\subsection{Testing the framework}
In this part, we compare predictions for mass functions, multiplicity functions and biases based on the Hertz and Stratonovich approximations with those calculated based on numerically generated excursion-set trajectories. For the bias parameters, one can use Eqs \eqref{lagrangian_bias4} and \eqref{bn0} for a given filter and barrier functions and calculate the parameters as summarized above. However, since our aim is to assess the performance of the Hertz and Sratonovich approximations by comparing their predictions with results from numerically generated large ensembles of excursion set trajectories (Monte Carlo-based results), it will be sufficient for our purpose to use the bias relation from Eq. \eqref{nu_PBS}. This also has the advantage of allowing us to generate Monte Carlo-based results for comparisons much more efficiently. However, in general, for non-Markovian trajectories, the multiplicity function is not guaranteed to be universal. Nevertheless, Eq. \eqref{nu_PBS} should provide a reasonable approximation for a broad class of barriers. This is particularly likely to be the case for barriers for which the quantity $B(S)/\sqrt{S}$, on which the FCD depends strongly, can be expressed entirely as a function of $\nu = \delta_c/\sqrt{S}$. 

One example is the constant barrier $B = \delta_c$. Another example is the family of SMT barriers, which can be written in the general form (\cite{2001MNRAS.323....1S}):
\begin{equation}
	\label{SMT_barrier}
	B(S, \delta_c) = \sqrt{a}\delta_c \left(1 + \beta(\frac{S}{a \delta_c^2})^\gamma \right),
\end{equation}
where $\delta_c$ is usually treated as a fixed value, while  $a$, $\beta$ and $\gamma$ are model parameters that can be fitted to simulation data. For the FCDs obtained with these classes of barriers, one can obtain $f(\nu)$ from Eq. \eqref{fs_to_fnu} and approximate the bias parameters from Eq. \eqref{nu_PBS}. Ultimately, one can obtain Eulerian bias parameters using proper transformations. Here, for comparisons, we focus on the first order Eulerian bias parameter which is obtained from: $b^{(\mathrm{E})}_1 = 1 + b^{(\mathrm{L})}_1$. The reason we focus on the first order bias is that, besides being the most important contribution at large scales, it is also more suitable for comparisons with our Monte Carlo-based results. This is because even a small numerical fluctuation in the Monte-Carlo based multiplicity functions can get significantly amplified upon differentiation, making comparisons with our predictions for higher order bias terms difficult. 

We consider both a constant barrier and the SMT barriers given by Eq. \eqref{SMT_barrier}, using different values of the parameters $a, \beta, \gamma$. For the smoothing filters, we consider Top-Hat, Gaussian and $W_2$ (\cite{2021MNRAS.508.3634M}) filters, which are defined in Fourier space as 
\begin{align}
	\label{filters1}
	&W_{\text{TH}}(x) = \frac{3}{x^3} (\sin{x} - x \cos{x}) = \frac{3}{x} j_1(x), \\\label{filters2}
	&W_{\text{G}}(x) = \exp\left(- \frac{x^2}{2} \right), \\\label{filters3}
	& W_2(x) = \frac{15}{x^2} j_2(x),
\end{align}
where $j_n$ denotes the spherical Bessel function of order $n$ and $x = k R$, with $R$ being the Lagrangian smoothing scale. 

For the numerical trajectory-based calculation of the FCD for a given smoothing filter and barrier function, we first generate logarithmically spaced values of $R$ and then calculate the corresponding $S$ values using the chosen filter function. We generate linear matter power spectra using the \texttt{Core Cosmology Library (CCL)}\footnote{https://github.com/LSSTDESC/CCL} (\cite{2019ApJS..242....2C}). The $S$ values are calculated over a suitable range, with a resolution sufficient for our comparisons. We then generate a large ensemble of  excursion set trajectories ($2\times10^9$ for this work) in the $S-\delta_S $ plane using the Cholesky method (\cite{2018MNRAS.478.5296N}). Next, we identify the locations of the first up-crossings of the trajectories using an appropriate interpolation method to improve the accuracy of the crossing point determinations. After identifying the first up-crossing points using the initial large set of $S$ values (densely sampled), we consider a new, smaller, set of $S$ values to calculate the FCD based on the first up-crossing points of the large set. For instance, for a given bin, $[S_i, S_{i+1})$, the corresponding FCD value is calculated as
\begin{equation}
	\label{fsi}
     f({S_{\text{mid}, i}}) = \frac{N_{\text{cross}}}{N_\text{total} \Delta S},
 \end{equation}
  where $S_{\text{mid}, i} = (S_i + S_{i+1})/2$, $N_{\text{cross}}$ is the number of trajectories that have up-crossed the barrier for the first time within that bin, $N_\text{total}$ is the total number of trajectories and $\Delta S = S_{i+1} - S_{i}$. The reason we use the midpoint, $S_{\text{mid}, i}$, as the point at which the corresponding FCD is evaluated is that, by definition, $N_{\text{cross}}/N_\text{total}$ is equal to the integral of $f(S)$ over the bin interval and, therefore, $f({S_{\text{mid}, i}}) \Delta S$ provides a better approximation for this integral. Also for comparisons in this work, we consider the following cosmological parameters for a flat $\Lambda$CDM universe: $\Omega_c = 0.25$,  $\Omega_b = 0.05$, $h = 0.68$, $n_s = 0.97$ and $\sigma_8 = 0.82$. 

  We first compare halo mass functions obtained from Eq. \eqref{mass_function} using  predicted and numerically computed (trajectory-based) FCDs. Fig. \ref{fig:fscmpr} shows the halo mass functions predicted from the Stratonovich approximation for the FCD, compared with their counterparts obtained based on numerically computed FCDs from Cholesky-generated trajectories. For this plot, we consider a top-hat filter and an SMT barrier, at two redshifts. Since the filter is top-hat in real space, the halo masses can be calculated from $M = \bar{\rho} V$, where $V$ is the volume of a sphere with Lagrangian radius $R$. One of the advantages of the Stratonovich approximation over more recent approximations for the FCD (e.g \cite{2024JCAP...10..079V}) is that it maintains consistently high accuracy in the large $S$ (low mass) regime with relative errors at the sub-percent or $\mathcal{O}(1\%)$ levels.
  
  Fig. \ref{fig:Bc} shows the multiplicity functions and biases obtained based on Hertz and Stratonovich approximations, as well as those calculated based on numerically generated excursion set trajectories using the three different smoothing filters defined in Eqs \eqref{filters1}--\eqref{filters3}, for a constant barrier $B = \delta_c = 1.686$. For comparison, the figure also shows the multiplicity function and bias obtained from a Markovian model. For the Markovian multiplicity function, we use the result based on the Press--Schechter model (\cite{1974ApJ...187..425P}),
  \begin{equation}
  	\label{nufnupsh}
  	\nu f_{\text{PS}}(\nu) = \sqrt{\frac{2}{\pi}} \nu \exp\left({- \frac{\nu^2}{ 2}}\right)
  \end{equation}
  and for the Markovian bias, we use the Mo and White result (\cite{1996MNRAS.282..347M}),
    \begin{equation}
    	\label{bMmw}
  	b_{\text{MW}} = 1 + \frac{\nu^2 - 1}{\delta_c}.
  \end{equation}
It is clear from this figure that non-Markovian effects become increasingly important as $\nu$ decreases (or, equivalently, as $S$ increases, corresponding to smaller halo mass scales), and that the Stratonovich approximation provides the best agreement with the numerical results in capturing such effects. Figs \eqref{fig:SMT756} and \eqref{fig:SM434_178} show the results for SMT barriers with different values of the parameters $a$, $\beta$ and $\gamma$. Again, one can see that the Stratonovich approximation provides the best agreement with the numerical predictions across all filter functions and barrier parameter values. 
 
Regarding the comparison between the Hertz and Stratonovich approximations across all figures, the performance of the Hertz approximation deteriorates as $\nu$ decreases. This is expected because smaller values of $\nu$ correspond to the larger-$S$ regime, where the probability of finding trajectories with multiple up-crossings increases. Following the arguments in Section \ref{sec:fs_hertz_stra}, the Hertz approximation is expected to perform worse than the Stratonovich approximation in correcting for the contributions of such multiple up-crossings. Regarding the Stratonovich approximation itself, we can see that it performs better at small $\nu$ for SMT barriers with larger values of the parameters $a$, $\beta$ and $\gamma$. This is because the probability of multiple up-crossings becomes smaller with such barriers compared to the constant barrier or SMT barriers with smaller parameter values. However, overall, the results across all figures indicate that the Stratonovich approximation performs best based on comparisons with the numerical results and captures the non-Markovian effects most accurately. 

\section{Conclusions}
\label{sec:conclusions}
In this work, we presented an efficient method for calculating halo multiplicity functions and bias parameters based on the previously presented Hertz and Stratonovich approximations for the first crossing distribution within the excursion set framework. The procedure for calculating the first crossing distribution for a generic barrier and smoothing filter is summarized in Subsection \ref{subsection:summary_fs}. In Section \ref{sec:multiplicity_bias_hertz_stra}, we discussed how halo bias parameters can, in general, be obtained in a numerically efficient manner using the framework presented in this work. We then showed examples of halo mass functions, multiplicity functions and biases calculated based on the Hertz and Stratonovich approximations, compared with numerical trajectory-based results in Figs \ref{fig:fscmpr}--\ref{fig:SM434_178}. Fig. \ref{fig:fscmpr} (mass function comparisons) corresponds specifically to the Stratonovich approximation. The multiplicity function and bias comparisons considered three different smoothing filters, together with constant and SMT moving barriers with different parameter values. In all cases, the Stratonovich approximation showed the best agreement with the numerical results. Furthermore, the Stratonovich approximation  maintains consistently high accuracy even at large values of the parameter $S$ (corresponding to the low-mass halo regime)  where non-Markovian effects become increasingly important.

Given the accuracy of the  Stratonovich approximation for calculation of the first crossing distribution, as well as the flexibility and computational efficiency of the framework presented in this work, this framework could provide an efficient tool for training differentiable emulators of halo abundances and bias parameters for use in cosmological analyses. Developing such emulators can be an interesting avenue for future work.

\section*{Acknowledgments}
The authors acknowledge the use of AI-based tools, including ChatGPT (OpenAI) and Claude (Anthropic), for assistance with improving the clarity and readability of the manuscript. SB is partially supported by the Abdus Salam International Center for Theoretical Physics (ICTP) under the regular associateship scheme. Moreover, SB is partially supported by the Sharif University of Technology Office of Vice President for Research under Grant No. QB050522.

\section*{Data Availability}
No external datasets were used in this work. All codes developed for the implementation of the framework are available upon request.



\bibliographystyle{mnras}
\bibliography{bias_refs} 





\bsp	
\label{lastpage}
\end{document}